\documentclass[conference]{IEEEtran}
\usepackage[pass]{geometry}

\usepackage{setspace}
\usepackage{graphicx}
\usepackage{subfigure}
\usepackage{float}
\usepackage{cite}

\usepackage{color}
\definecolor{darkblue}{rgb}{0,0,.75}
\usepackage{listings} 
\lstnewenvironment{PseudoCode}[1][]
{\lstset{language=Matlab,basicstyle=\scriptsize, keywordstyle=\color{darkblue},numbers=left,xleftmargin=.04\textwidth,#1}}
{}

\begin{document}

\title{Improving the scalabiliy of neutron cross-section lookup codes on multicore NUMA system}


\author{
\IEEEauthorblockN{Kazutomo Yoshii, John Tramm, Andrew Siegel, Pete Beckman}
\vspace{5pt}
\IEEEauthorblockA{
Argonne National Laboratory\\ 
Mathematics and Computer Science Division\\
9700 South Cass Avenue, Argonne, IL}  
}

\maketitle

\begin{abstract}

We use the XSBench proxy application, a memory-intensive OpenMP
program, to explore the source of on-node scalability degradation of a
popular Monte Carlo (MC) reactor physics benchmark on non-uniform memory
access (NUMA) systems.  As background, we present the details of
XSBench, a performance abstraction ``proxy app'' for the full MC
simulation, as well as the internal design of the Linux kernel.  We
explain how the physical memory allocation inside the kernel affects
the multicore scalability of XSBench.  On a sixteen-core, two-socket
NUMA testbed, the scaling efficiency is
improved from a nonoptimized 70\% to an optimized 95\%, and the
optimized version consumes 25\% less energy than does the nonoptimized
version.  In addition to the NUMA optimization we evaluate a
page-size optimization to XSBench and observe a 1.5x performance
improvement, compared with a nonoptimized one.

\end{abstract}


\section{Introduction}
\label{intro}

Improved processor performance on the path
to exascale computing~\cite{darpa_exascale,iesp-roadmap} is widely expected to involve increasing degrees of on-node
concurrency. This hardware trajectory in turn has deep implications
for the adoption of mathematical models, specific choice of discretization,
and implementation on next-generation compute resources. 

An excellent example involves the choice between deterministic and
Monte Carlo (MC) approaches to particle transport. Since MC
methods fundamentally are embarrassingly parallel over particle tracks,
for many classes of MC problems the expected performance improvements
are likely to enable calculations that were once impractical because of
unrealistically long integration times.  This situation is true especially for
classic reactor physics calculations, where robust computations of
detailed neutron distributions in a realistic reactor core are largely
still beyond the reach of current compute capabilities. As we begin to
take advantage of new multicore architectures, however, the picture
is starting to change. Indeed, contrary to initial expectations,
new challenges are arising. For example, although MC methods  
are naturally parallel, we have already observed scalability
limits~\cite{tramm2013memory} even on today's
multicore processors, particularly with non-uniform memory access
(NUMA) systems.  Therefore, identifying the specific causes of
these scalability bottlenecks is urgent. 

Typically,
researchers have relied on processors' performance counters
to identify bottlenecks in high-performance computing (HPC) applications. 
The counter number itself, however, is no longer sufficient for analyzing
scalability problems in a modern compute node because of operating system
(OS) abstractions.  The OS abstraction is an important concept
for portability, but it creates a gap between the hardware and
the application. With multicore scaling, how OS subsystems
such as the task scheduler and memory management manage hardware resources
becomes key. The performance counters cannot tell us the OS
behavior directly.  The OS scheduler may load-balance processes
automatically in order to maximize hardware utilization unless explicitly
specified, thus affecting cache locality.  The OS memory management
basically hides physical memory allocation from applications, but it
affects the performance if memory access latency is nonuniform.
Additionally, a runtime system such as OpenMP~\cite{openmp-oarb} does
dynamic balancing, with almost no coordination between the OS and
runtime. Thus, cross-cutting analysis is becoming vital in order to
understand the interaction between each layer and to identify the real
cause of the scalability problems of MC methods on multicore.

This paper focuses on popular modern multicore NUMA processors such as
Intel Xeon SandyBridge processors~\cite{10.1109/MM.2012.12}. 
More than 80\% of the processors in the TOP500 list are
Intel Xeon processors, which are multicore processors with symmetric
multithreading.  Xeon systems are usually multi socket with
NUMA, which is an architectural
design to mitigate the memory bottleneck by adding additional memory
controllers to a core or a set of cores.  Bigger systems may have more
than 100 cores; 15-core Intel Ivy Bridge processors can scale up to 8 sockets (120 cores in total).
With NUMA, memory access latency is not
uniform. Thus, exploiting memory locality is an important
factor in achieving good performance. This factor, however, poses a big
challenge to both the system software and applications.  


In the present analysis we focus on the Linux kernel,
the de facto standard OS kernel in high-performance computing. The Linux environment offers
numerous benefits to users.  One benefit is that myriad tools
and libraries are available in the Linux environment, such as the GNU
debugger, strace, valgrind, and LLVM.
These
software tools not only are powerful but also are continuously
debugged and updated because Linux or open source communities attract
numerous users and developers.  According to the TOP500~\cite{top500}
statistics in June 2014, 485 of the 500 entries (97\%) are categorized
as belonging to the Linux family.

Technically a few of those listed are not true
Linux kernels. IBM's Blue Gene architecture, for example, runs
CNK~\cite{Giampapa:2010:ELS:1884643.1884667}, which is a lightweight
OS kernel optimized for HPC applications written from scratch by IBM;
but it supports Linux application binary interface and a subset
of Linux system calls (multitasking related system calls are not
supported, however). CNK is designed to yield the real hardware
performance to applications with statically mapping memory.  Because of
its overly simplified memory management and task scheduler, however, multitask-related
system calls are not implemented or partially implemented in a limited way, such
as \texttt{clone()} and \texttt{mprotect()}; thus CNK users cannot benefit from some of the Linux
tools such as strace.  

Unlike CNK, Linux kernels are primarily optimized for handle desktop and/or server
workloads that consist of various tasks that could be
short lived, CPU intensive, I/O intensive, highly interactive, or a
combination of all of these.  Most tasks are loosely coupled or
independent.  Handling them fairly and efficiently is 
challenging, and the current Linux subsystems tend to cause negative
effects on parallel HPC applications (e.g.,OpenMP, MPI) that
monopolize virtually all node resources.  The Linux kernel treats
those MPI or OpenMP applications as a collection of regular processes
or threads and schedules them without any bias.  The
Linux kernel provides several transparent mechanisms to optimize 
the performance such as process migration, NUMA balancing, transparent hugetlbfs.
However, these transparent mechanisms may also wider the gap between the hardware and 
the application and complicate the resource identification in user space.

Our contributions include the following:

\begin{itemize}

\item Understanding of the multicore scalability problem on NUMA,
  analyzing both details XSBench and the Linux kernel design.


\item Analysis of the importance of physical memory allocation in the Linux
  kernel. We present a few user-space NUMA optimizations to mitigate the scalability
  problem and improve the performance.


\item Detailed scalability and performance analysis of the NUMA optimizations, including
  energy consumption, on different running modes.

\end{itemize}

\section{The XSBench Proxy Application}
\label{sec:xebench}

Monte Carlo methods of reactor simulation have a prohibitively
long time to solution on current-generation supercomputers, although
the embarrassingly parallel nature of the MC particle transport
algorithm suggests that it should be an exceptional candidate for
good performance scaling on exascale class supercomputers. Thus,
exascale supercomputers offer the possibility of completing a robust,
full-core nuclear reactor simulation with hundreds of
nuclides and millions of geometric regions within a reasonable
wall time, opening new avenues in reactor design.  Recent
studies~\cite{tramm2013memory, Siegel:2013vn}, however, have found that the
MC transport algorithm is generally bound by bandwidth and DRAM
latency, rather than by the floating-point capabilities of modern
processors. Since the likely path to exascale will involve
greatly increasing the floating-point capacities of nodes, while
only marginally increasing the bandwidth~\cite{Dosanjh2013,
Attig2011, Rajovic2013, Engelmann2013}, it is extremely important
to optimize the application, software runtime, OS, and hardware
in order to maximize bandwidth efficiency.

To this end, the XSBench proxy application was created. It abstracts
the key performance aspects of full-scale MC transport codes, such
as OpenMC~\cite{Romano:2013hj} and MCNP~\cite{MCNP}, into a smaller
package that is easier to port, run, and analyze on various 
novel and experimental architectures. XSBench executes only macroscopic
neutron cross-section lookups, a key computational kernel
in MC transport applications that constitutes 85\% of the total runtime
of OpenMC~\cite{tramm2013memory}.  XSBench has been shown to
accurately mimic the computational requirements of full-scale MC
transport applications~\cite{Tramm:wy}, so performance analysis done with
XSBench will translate well to full-scale applications.  XSBench
is written in C, with node-level parallelism support by OpenMP.
Reactor parameters that define the size and scope of the problem,
such as the number of nuclides and materials used, are based on a
well-known community reactor benchmark model, the Hoogenboom-Martin
model~\cite{Hoogenboom:2010}. XSBench is developed by the Center
for Exascale Simulation of Advanced Reactors (CESAR) and is an open
source software project~\cite{xsbench:Online}.

\subsection{Algorithm and Data Structure}
\label{sec:xsalgo}

\begin{figure} 
\centering
\includegraphics[width=.7\columnwidth]{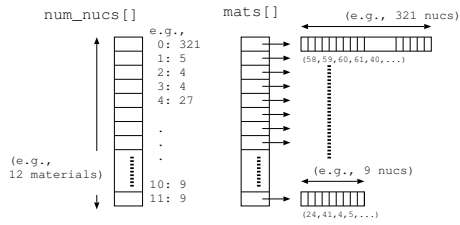}
\caption{Material}
\label{fig:material}
\end{figure}

\begin{figure} 
\centering
\includegraphics[width=.7\columnwidth]{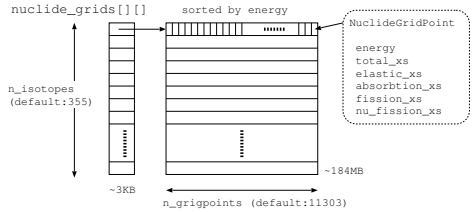}
\caption{Nuclide grids}
\label{fig:nuclidegrids}
\end{figure}

Figure~\ref{fig:pseudobasic} is a C-like pseudo code that presents a basic
idea of the MC particle transport algorithm.  The major data structures
involved in the basic algorithm are the material data and the nuclide grids data 
shown in Figure~\ref{fig:material} and
Figure~\ref{fig:nuclidegrids}. 
With default runtime configuration (355 nuclides
tracked, roughly 4 million gridpoints, and 5 cross-section interaction
types), the nuclide grids data consumes approximately 184 MB.
Compared with the nuclide grids data, the material data size is
negligible.  Each iteration is highly independent, one can
easily exploit node-level parallelism using the OpenMP
parallel for loop.  However, the computational cost of the lookup is
quasilinear (see Table~\ref{tbl:cost}) because of a binary search in the
innermost loop.

\begin{figure} 
\begin{PseudoCode}
parallel for :
  p_energy = pick energy randomly 
  p_mat    = pick material randomly
  xs_vector = 0
  for j in num_nucs[p_mat] :
     nuc = mats[p_mat][j]   // get nuclide ID
     pos = binary search p_energy in nulide_grids[nuc]
     xs_vector += interpolate nuclide_grids[nuc] at pos
  (output: xs_vector)
\end{PseudoCode}
\caption{Pseudo code: cross-section lookup (basic algorithm)}
\label{fig:pseudobasic}
\end{figure}

\begin{figure} 
\centering
\includegraphics[width=.6\columnwidth]{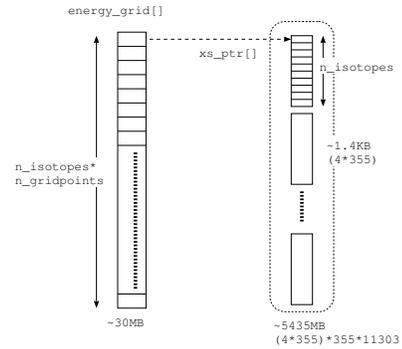}
\caption{Energy Grids}
\label{fig:energygrids}
\end{figure}

Using the unionized energy grid, described by
Lepp\"{a}nen~\cite{Leppanen:2009} and Romano~\cite{Romano:2013hj}, one can improve the computational cost of the
lookup to linear from quasilinear (see Table~\ref{tbl:cost}).
However, the drawback of the unionized energy grid is its memory footprint.
In this study we use the unionized energy grid.
Figure~\ref{fig:energygrids} presents the unionized energy grid
implemented in XSBench. This structure holds the sorted energy of all
nuclide grid points and the pre-calculated closest corresponding
energy level on each of the different nuclide grids. 

\begin{table}[htbp]
\caption{Lookup Algorithm Comparison}
\label{tbl:cost}
\footnotesize
\centering
\begin{tabular}{|l|l|l|}
\hline
&\\[-1em]
   & \textbf{Lookup Cost} & \textbf{Memory Requirement} \\ \hline
&\\[-1em]
Basic & quasilinear &  ~184\,MB \\ 
Unionized grid & linear &  ~5617\,MB \\ \hline
\end{tabular}
\end{table}

The algorithm exhibits highly random memory access patterns due to
multiple levels of indirect memory accesses.  With the default
configuration, nuclide\_grids is 184\,MB in size, which fits only in
the last level of cache and is likely to cause translation lookaside
buffer (TLB) misses. On the Intel Xeon node, every access to this
structure causes a TLB miss with the default 4\,KB page size because
the Intel Xeon has only 64 TLB entries.

\begin{figure} 
\begin{PseudoCode}
parallel for :
  p_energy = pick energy randomly 
  p_mat    = pick material randomly
  idx = binary search energy on energy_grid
  xs_vector = 0
  for j in num_nucs[p_mat] :
     nuc = mats[p_mat][j]   // get nuclide ID
     pos = energy_grid[idx].xs_ptrs[nuc]
     xs_vector += interpolate nuclide_grids[nuc] at pos
  (output: xs_vector)
\end{PseudoCode}
\caption{Pseudo code: cross-section lookup (with unionized energy grids)}
\label{fig:pseudo}
\end{figure}

\subsection{Multicore Scalability}
\label{sec:scalability}

\begin{figure} 
  \centering
  \includegraphics[width=.6\columnwidth]{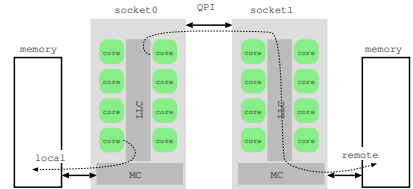}
  \caption{A dual socket Sandy Bridge NUMA node}
  \label{fig:dualsocket}
\end{figure}

To observe the multicore scaling efficiency, we run XSBench on an
Intel Sandy Bridge node, changing the number of OpenMP threads.  The
Sandy Bridge node includes two Xeon E5-2670 processors, 8 cores and 16
hardware threads with hyper-threading (HT), which runs at 2.6\,GHz, up
to 3.3\,GHz with the Intel turboboost technology.  These two
processors are connected via two Intel Quick Path Interconnect (QPI)
links, which forms a cache-coherent NUMA node (see
Figure~\ref{fig:dualsocket}). The node runs Linux kernel version 3.16
with NUMA enabled.

\begin{figure}
  \centering
  \includegraphics[width=.6\columnwidth]{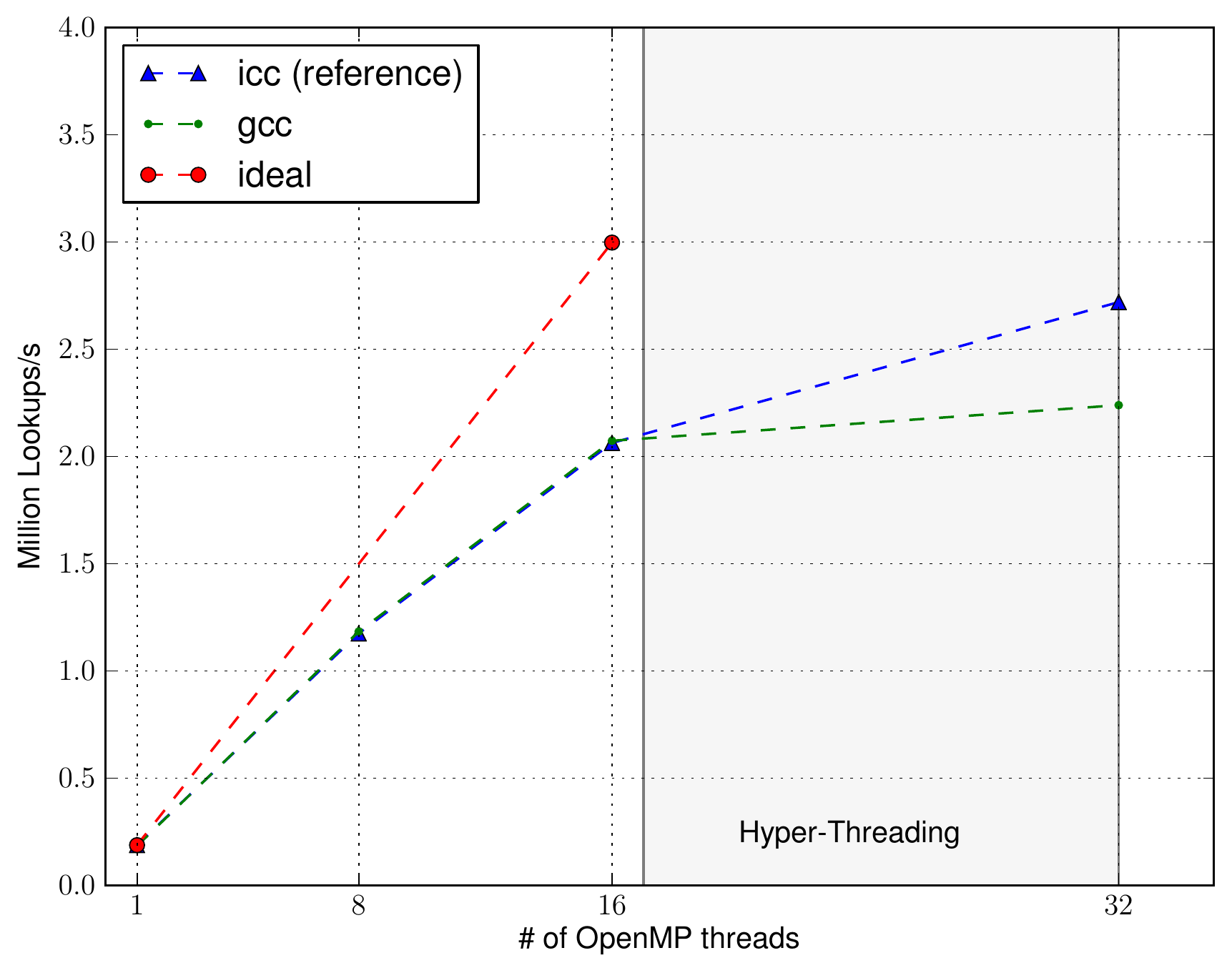}
  \caption{Cross section lookup}
\label{fig:xsperf}
\end{figure}

\begin{figure}
  \centering
  \includegraphics[width=.6\columnwidth]{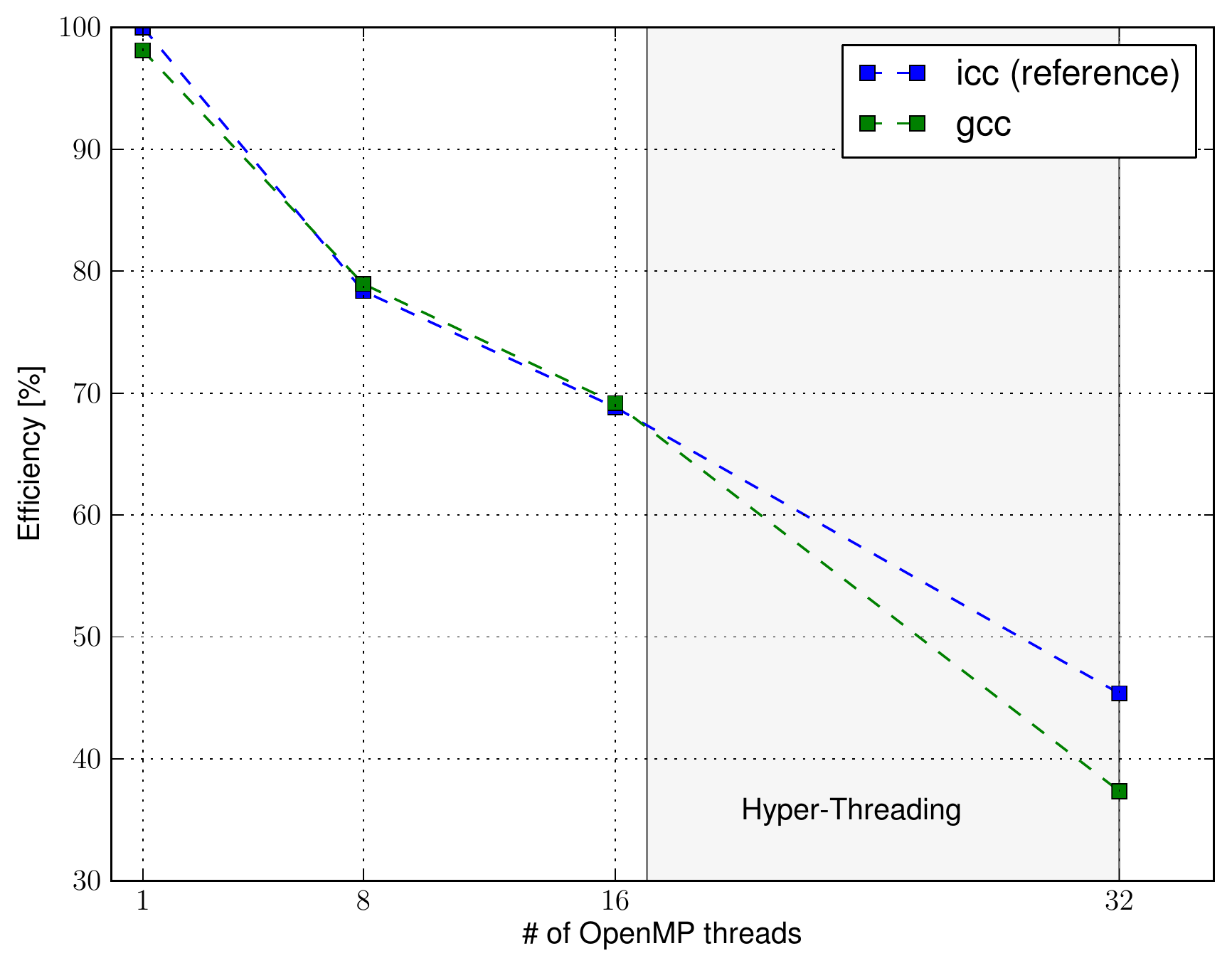}
  \caption{Multicore scaling}
\label{fig:xsscaling}
\end{figure}

Figure~\ref{fig:xsperf} is the cross-section lookup
performance, measuring XSBench run at 2.6\,GHz.
Figure~\ref{fig:xsscaling} is the multicore scalability efficiency,
which is calculated by the following equation.

\begin{equation}
Efficiency_n (\%) = \frac{P_n}{P_1 * n} * 100
\end{equation}

where n is the number of threads and $P_n$ is the measured
performance.  The scaling efficiency is dropped to approximately
70\,\% on 16 OpenMP threads.  This result matches with the scaling
problem previously reported by Tramm et al.~\cite{tramm2013memory} on
an Intel Xeon NUMA node, two eight-core E5-2650 processors runs at
2\,GHz and up to 2.8\,GHz with turboboost. They observed approximately
1.6 million lookups/s and 70\,\% efficiency at 16 threads. However on
a IBM BlueGene/Q (BGQ) node, the scaling efficiency is only dropped to
96\,\% at 16 threads. The major difference between the BGQ node and
the Xeon node is that the BGQ node is a uniform memory access shared
memory architecture while the Xeon nodes are NUMA.  Other notable
difference is that the BGQ node runs a custom OS kernel that is
optimized for HPC workloads while the Xeon nodes runs the Linux
kernel.  In the previous study only marginal performance gains are
observed with HT. We find that XSBench compiled by an Intel compiler
(icc) shows a reasonable performance gain with HT.

%


\section{Operating System and Runtime System}
\label{sec:os}

Data locality is a critical factor for performance on a NUMA node.
However, controlling NUMA-aware data placement is not a simple task for
user-space codes because of the OS abstraction, the virtual address,
and load balancing.  In this section we detail the Linux kernel's
internal design and explain why the current design affects the
multicore scalability of a memory-intensive OpenMP parallel code on a
NUMA node.

\subsection{Kernel Management Structure and OpenMP}
\label{sec:osstruct}

\begin{figure} 
\centering
\includegraphics[width=.8\columnwidth]{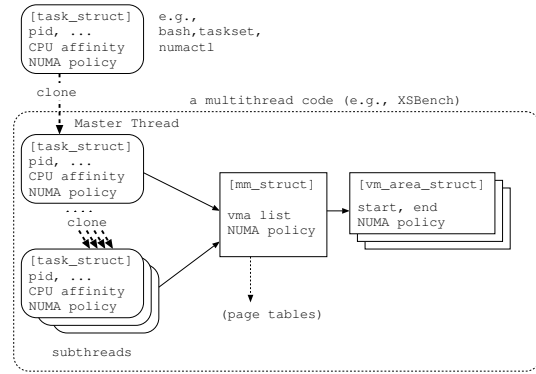}
\caption{Kernel management structure}
\label{fig:kernelstruct}
\end{figure}

Figure~\ref{fig:kernelstruct} is a simplified view of the major
management structure in the Linux kernel related to this study.  In
this figure, each rounded rectangle represents a process or thread, and
each (shape-edged) rectangle represents an internal data structure.
The ``task\_struct'' structure contains all the information related to
a process or thread such as process ID.  The difference between
process and thread is that a process has its own ``mm\_struct'' while
a thread shares the parent's ``mm\_struct.'' An OpenMP program, for
example, is simply a collection of a process (the master thread) and
one or more threads that are cloned by the master thread. Thus the
Linux kernel does not distinguish an OpenMP program from others.

When a new process/thread is created, many attributes are inherited
by the child process 
from a parent, including the CPU affinity mask and
the NUMA policy.  In fact, the \texttt{taskset} command leverages this
behavior in order to control the child process's CPU affinity mask (e.g., start
a program on the second core). However, the \texttt{taskset} command
is unable to control the affinity mask of non-master threads in an
OpenMP program because the attribute inheritance is only from its
parent process.  For a multithreaded program, each thread needs to
call the \texttt{sched\_setaffinity()} system call in order to bind an
OpenMP thread to a specific CPU core.

\subsection{Physical Memory Allocation}

One of the important roles of the OS kernel is to provide a linear
virtual address space to each process/thread while keeping better
physical memory utilization.  The Linux kernel, for example, relies on
a memory management unit (MMU) that supports paged virtual memory, which
divides a virtual address space into contiguous memory blocks called
pages.  With a normal operating mode (e.g., long mode on x86-64), CPU
instructions cannot access data with physical addresses directly, technically even
inside the OS kernel because processors always translate a virtual
address to a physical address.  The OS kernel is responsible for
managing in-memory page tables for virtual-to-physical address
translation. Processors normally have a translation lookaside buffer
to cache the content of recent page tables, because scanning
through in-memory page tables for every translation is prohibitively
expensive.  In general TLB is a scarce resource; the number of TLB
entries is limited, so TLB misses still occur, leading to
performance
degradation~\cite{Chen:1992:SBS:146628.139708,yoshii:2010:ijhpca}.  A
larger page size\footnote{The start address of each page must be
  aligned with its own page size.} can reduce TLB misses and improve
the memory access performance if
available~\cite{lwn_gorman_huge_pages}. Table~\ref{tbl:snbtlb} shows TLB sizes and the
number of the TLB slots on available page sizes in the Intel Sandy
Bridge microarchitecture. With the 2\,MB page, for example, a 64\,MB
of memory area can be accessed without TLB misses.

\begin{table}
\caption{Intel Sandy Bridge's data TLB}
\label{tbl:snbtlb}
\footnotesize
\centering
\begin{tabular}{|l|l|l|l|}
\hline
                  & \textbf{4\,KB} & \textbf{2\,MB} & \textbf{1\,GB}  \\ \hline
First level   & 64     & 32          & 4 \\ \hline
Second level & 512    & node        & none \\ \hline
\end{tabular}
\end{table}

\begin{figure} 
\centering
\includegraphics[width=.8\columnwidth]{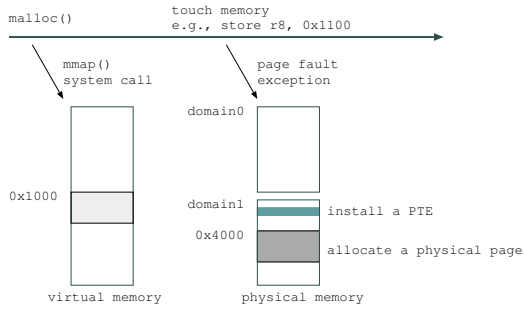}
\caption{Memory allocation}
\label{fig:memoryallocation}
\end{figure}

Figure~\ref{fig:memoryallocation} is an example of the flow of a
typical memory allocation in the Linux environment. First an
application process calls a memory allocation function such as the
\texttt{malloc()} library function, which internally invokes the 
\texttt{mmap()} system call.  The \texttt{mmap()} itself only creates
a virtual memory area when it is invoked; physical memory is not
allocated at this point unless the MAP\_POPULATE flag is specified.
In Figure~\ref{fig:kernelstruct}, the \texttt{mmap()} manipulates
``vm\_area\_struct.''  When the user-space program attempts to store to
or touch the virtual memory range created by \texttt{mmap()} for the
first time, the processor raises a page fault exception, and the Linux
kernel allocates a physical memory page and installs a page table entry
into main memory for virtual-to-physical address translation.

\subsection{NUMA Memory Domain}
\label{sec:osnuma}

\begin{figure} 
\centering
\includegraphics[width=.6\columnwidth]{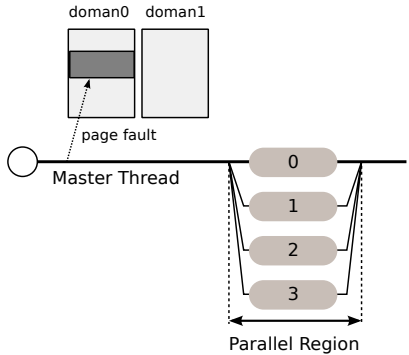}
\caption{A multithreaded program}
\label{fig:multithread}
\end{figure}

A NUMA system has multiple memory domains. Our Sandy Bridge testbed,
for example, has two NUMA domains. By default, the Linux kernel
allocates a physical page from the NUMA domain where a thread causes
the page fault. This is called ``first-touch.''  XSBench, an OpenMP
multithreaded application, allocates and touches the data buffer from
the master thread (Figure~\ref{fig:multithread}). By default, a
physical page is allocated from the domain where the master thread
causes a page fault; hence, the application's data buffer is
likely to be in one of the NUMA domains.  During its computational
phase in the parallel region (Figure~\ref{fig:multithread}) that
spreads over all domains, half of the CPU cores have to incur
expensive remote memory accesses.  This is a typical cause of the
multicore scalability problem on a NUMA node, particularly for a
memory-intensive OpenMP code such as XSBench. The
problem is expected to become much more pronounced as the number of the NUMA domains
increases and the system's energy efficiency is drastically
reduced.

The Linux kernel detects the NUMA topology of the system during the
boot time (e.g., by parsing the system resource affinity table on the
Sandy Bridge system) and initializes the NUMA scheduler domains with
the detected topology.  The Linux kernel internally has service
routines to identify the NUMA domain ID associated with the CPU or the
memory range of each NUMA domain. The Linux kernel exposes the NUMA
topology information to the user space via the sysfs virtual file system,
but user-space programs are unable to identify the NUMA domain ID
related to a specific memory address. Although one can 
look up a physical address from a process's virtual address using the
pagemap interface, the Linux kernel currently does not provide a
mechanism for the user space to look up the NUMA domain ID related to the
physical address.

\subsection{Load Balancing}

The Linux kernel distributes workloads over the CPU cores by migrating
processes/threads in order to utilize all the resources
efficiently. It takes into account various runtime attributes such as
CPU business/idleness, cache hotness, memory pressure, and scheduling
domains in order to choose a CPU core that a process runs, considering
the process's CPU affinity mask.\footnote{By default, a process's CPU affinity
  mask is set to all CPUs.}  In fact the Linux kernel scheduler is
aware of the NUMA domain, migrating processes within the associated
NUMA domain as much as possible.  With the previous XSBench example,
its master thread is likely to stay in the same NUMA domain.


The recent Linux kernel (e.g., 3.8 or later) has a configuration
option for automatic NUMA-aware memory placement, which utilizes the
page fault handling mechanism to detect expensive remote
accesses. Once these are detected, it attempts to migrate the physical pages
close to the thread that caused the page fault.

However, the process migration and the automatic NUMA balancing
basically conflict with each other, because locality is important to
the NUMA balancing while equal distribution is important to the CPU
load-balancing.  In general the Linux kernel is optimized primarily
for general-purpose workloads such as server workloads that consist
of independent, irregular tasks.  On the other hand, HPC workloads
are usually parallel and tend to monopolize the resources for a long
duration; thus such automatic or transparent mechanisms tend to have
a negative effect on HPC workloads.

\subsection{NUMA Memory Policy}
\label{sec:numapolicy}

\begin{figure} 
  \centering
  \subfigure[Bind]{\includegraphics[width=.3\columnwidth]{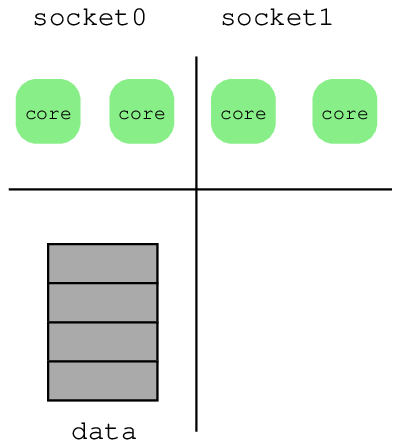}}\quad
  \subfigure[Interleaved]{\includegraphics[width=.3\columnwidth]{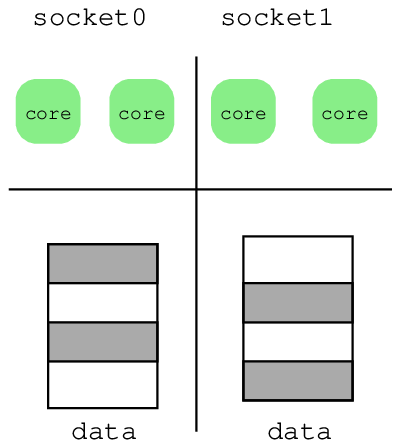}}\quad
  \caption{NUMA memory policy}
  \label{fig:dataplacement}
\end{figure}

The Linux kernel provides two system calls to control the NUMA
policy: the \texttt{set\_mempolicy()} system call sets the policy per
task, and the \texttt{mbind()} system call sets the policy of each
virtual memory range individually.  The major NUMA policy includes
default, bind, and interleaved (see Figure~\ref{fig:dataplacement}).
The default policy is the first-touch policy previously described.
The bind policy restricts physical memory allocation to specified
domains. The interleaved policy interleaves physical memory allocation
within specified nodes. The granularity of the interleave is the page
size (e.g., 4\,KB). The unionized energy grids would be a perfect candidate for
this policy.

The interleave policy can convert NUMA to a reasonable uniform memory
access  with a performance penalty, which potentially mitigates
the scalability problem but stresses the interconnect. 


\section{Memory Optimizations}
\label{sec:opt}

In the previous section we explained how an unbalanced distribution of the
physical memory allocation prevents XSBench from scaling on a
multicore NUMA system.  In this section we present three 
memory optimizations: ``numactl,'' ``numag,'' and ``numag+hugetlb.''

\subsubsection{default:}

The default provides no memory optimization.  XSBench is
executed with the default CPU affinity mask (set to all CPUs), which
means the Linux kernel can migrate XSBench threads within their
default scheduling policy.

\subsubsection{numactl:}

The numactl optimization requires no code modification.  XSBench is
executed via the \texttt{numactl} command with the
``--interleave=all'' option, which sets the interleave policy to all
XSBench threads, with the granularity of the default page size
(4\,KB). Internally the \texttt{numactl} command sets the interleave
flag to the NUMA policy attribute in ``task\_struct'' (see
Figure~\ref{fig:kernelstruct}) and starts XSBench.  The NUMA policy is
inherited from the XSBench master thread to its child threads; thus
all XSBench threads are set to the interleave.

\subsubsection{numag:}

The numag optimization requires a minimum code modification: the target
memory allocation functions needs to be replaced with custom
NUMA-aware memory allocation functions provided by a small library
called ``numag,'' which we implement for this experiment.  The major
functionality of the ``numag'' library is summarized as follows.

\begin{itemize}
\item Find a socket ID (a NUMA domain ID) from a CPU ID.
\item Find a per-socket master from a CPU ID.
\item Allocate a buffer in all sockets. This is used to duplicate
  read-only data.
\item Allocate a buffer with interleaved enabled, which calls the
  \texttt{mbind()} system call internally.
\end{itemize}

In order to provide this functionality, the numag library has to
disable the process migration by strictly binding OpenMP threads to
CPUs.

\begin{figure} 
  \centering
  \includegraphics[width=0.6\columnwidth]{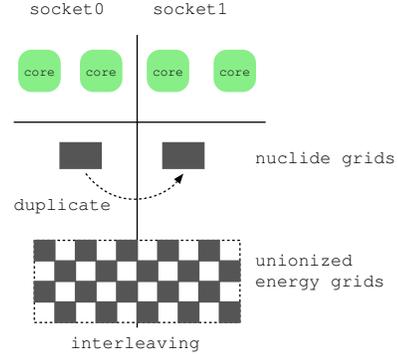}
  \caption{XSBench data placement}
  \label{fig:numag2}
\end{figure}

As for the XSBench data structures, the nuclide grids data structure
(Figure~\ref{fig:nuclidegrids}) is a good candidate for duplicating
because it is the most frequently accessed structure and the access
pattern to this structure is highly random.  With the data structure
size 184\,MB and the Sandy Bridge TLB
specification (Table~\ref{tbl:snbtlb}), every data access is
likely to end up with a TLB miss. If this structure is allocated on a
remote node, for example, page tables associated with this structure are
also located in the remote node.  Refilling TLB entries and loading actual
data remotely increase both the access latency and the energy
consumption.  The good news is that the data size is relatively small;
duplicating this structure in every socket will not create critical
memory pressure.
On the other hand, the unionized energy grids (see
Figure~\ref{fig:energygrids}) is huge and is less frequently
accessed than the nuclide grids. We decided to interleave the
unionized energy grids in the numag optimization.
Figure~\ref{fig:numag2} depicts how the numag optimization allocates
XSBench's data structures.

\subsubsection{numag+hugetlb:}

The numag library also provides a function that allocates a buffer
that is mapped with 2\,MB pages explicitly using the Linux kernel hugetlb
support. Note that the Linux kernel's transparent hugetlb support is
disabled in our test environment.  In this optimization we only
allocate the nuclide grids with 2\,MB pages; others are allocated
with the default 4\,KB pages.  This is technically not a NUMA
optimization; however, it does reduce the memory traffic regarding TLB
misses. 

\section{Results}
\label{sec:results}

We evaluate the memory optimizations on the Intel Sandy Bridge--based
dual-socket NUMA node (described in Section~\ref{sec:scalability}).
Here we first compare
the performance between the optimizations and shows that the
optimizations mitigate the multicore scalability
(Section~\ref{sec:numaopt}).
We then
compare the performance and energy
consumption between the optimizations on four different running modes
(Section~\ref{sec:runmode}).

\subsection{Memory Optimizations}
\label{sec:numaopt}

\begin{figure}
\centering
\includegraphics[width=.6\columnwidth]{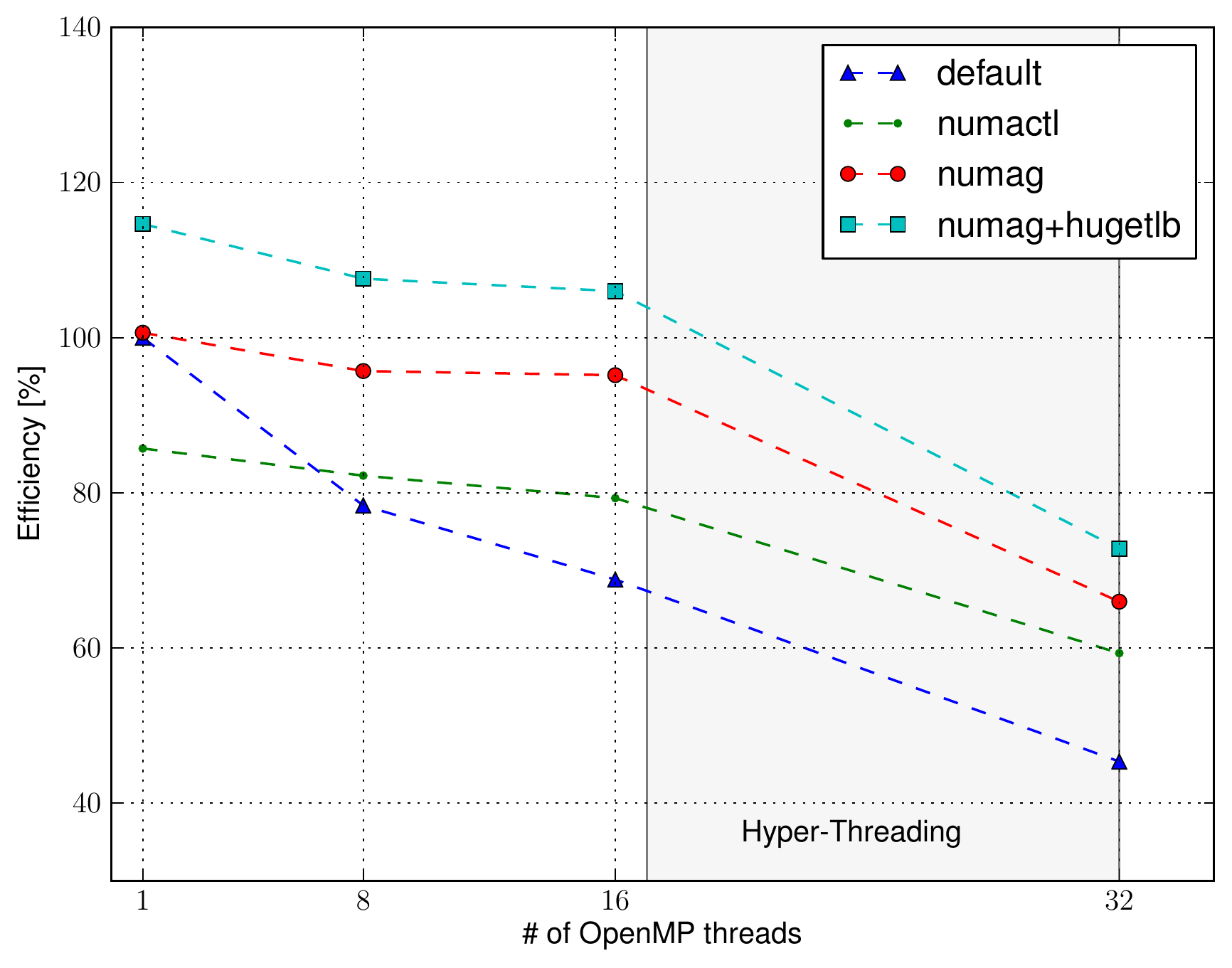}
\caption{Multicore scaling}
\label{fig:optxsscaling}
\end{figure}

Figure~\ref{fig:optxsscaling} shows a comparison of the multicore
scalability efficiency of XSBench with the three memory optimizations
described in the previous section; the running environment is the same as that 
used in Section~\ref{sec:scalability}.
The 
scaling efficiency in this comparison is calculated by the following
equation:

\begin{equation}
Efficiency_n (\%) = \frac{P_n}{Pd_1 * n} * 100,
\end{equation}

\noindent
where $n$ is the number of threads, $P_n$ is the measured performance,
and $Pd_1$ is the measured performance of ``default.''

The results clearly show that both ``numactl'' and ``numag'' improve
the multicore scalability. With one thread, however, the ``numactl''
performance improvement is less than that of the ``default.'' The reason is that the data is
interleaved, so half the data requires expensive remote access.  At
16 threads, the efficiency is improved to 80\% from a nonoptimized
70\% with ``numactl,'' which requires no code modification; and it is
improved to approximately 95\% with ``numag,'' which is close to that
of IBM BG/Q uniform memory access node observed by Tramm et al.~\cite{tramm2013memory}.
Although the efficiency drastically drops at 32 threads, which is in
hyper-threading (HT), this is expected because threads share the core
resources such as L1 cache.

\begin{figure}
\centering
\includegraphics[width=.6\columnwidth]{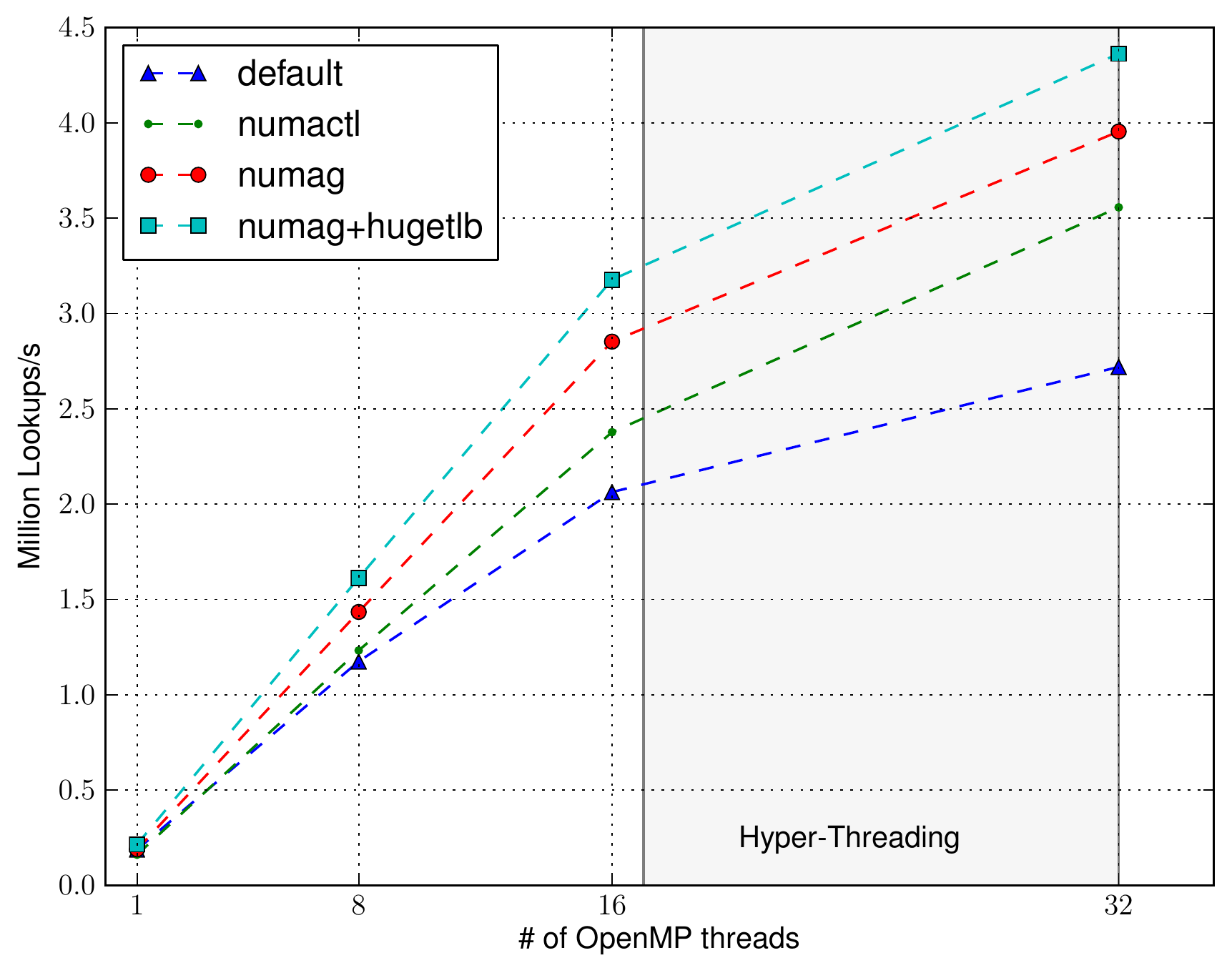}
\caption{Cross-section lookup}
\label{fig:optxsperf}
\end{figure}

Figure~\ref{fig:optxsperf} show a comparison of the neutron cross-section
lookup performance among the memory optimizations.  
We note that hyper-threading is effective for all memory
optimizations. With ``default,'' XSBench achieves 2.7
MLookups/s. Without any code modification (``numactl''), it achieves
3.6 MLookups/s. With the maximum optimization, it achieves 4.4
MLookups/s with hyper-threading, which is a considerable improvement.

%

\begin{figure}
\centering
\includegraphics[width=.6\columnwidth]{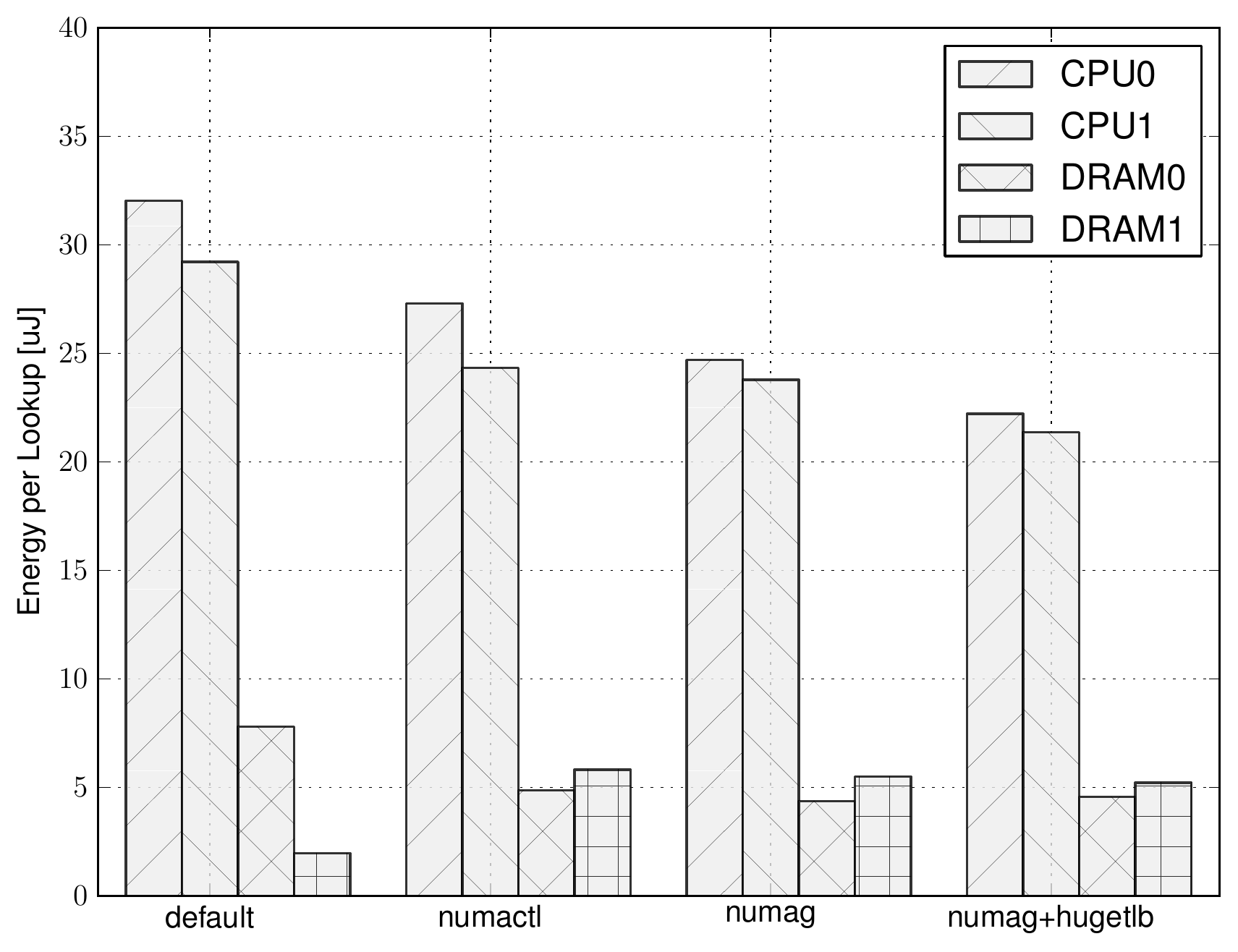}
\caption{Energy breakdown at 16 threads}
\label{fig:optxsbreak}
\end{figure}

In addition to the regular scalability and performance analysis, we
measure the detailed energy consumption of each memory
optimization (see Figure~\ref{fig:optxsbreak}) at 16 threads, by
reading Intel running average power limit (RAPL) counters through the
\texttt{sysfs} interface.\footnote{The newer Linux kernel provides the
  sysfs interface for both measuring energy and capping power using
  RAPL.}  CPU0 and CPU1 are an 8-core CPU in socket0 (domain0) and
socket1 (doamin1), respectively. DRAM0 and DRAM1 are CPU0's DRAM and
CPU1's DRAM, respectively.

With ``default,'' DRAM0 consumes more than DRAM1 does. The reason is that
XSBench's data are primarily allocated with socket0.  The DRAM energy
consumption is basically proportional to the total number of 
memory requests. With the NUMA optimizations, both DRAM0 and DRAM1
consume about the same energy because data accesses are balanced out.
Moreover, the total DRAM consumption is about the same.
On the other hand, the CPU energy consumption is basically proportional
to the time to solution; that is, the total CPU energy consumption decreases
with higher optimization. 



\subsection{Running Modes}
\label{sec:runmode}

\begin{figure}
\centering
  \includegraphics[width=.6\columnwidth]{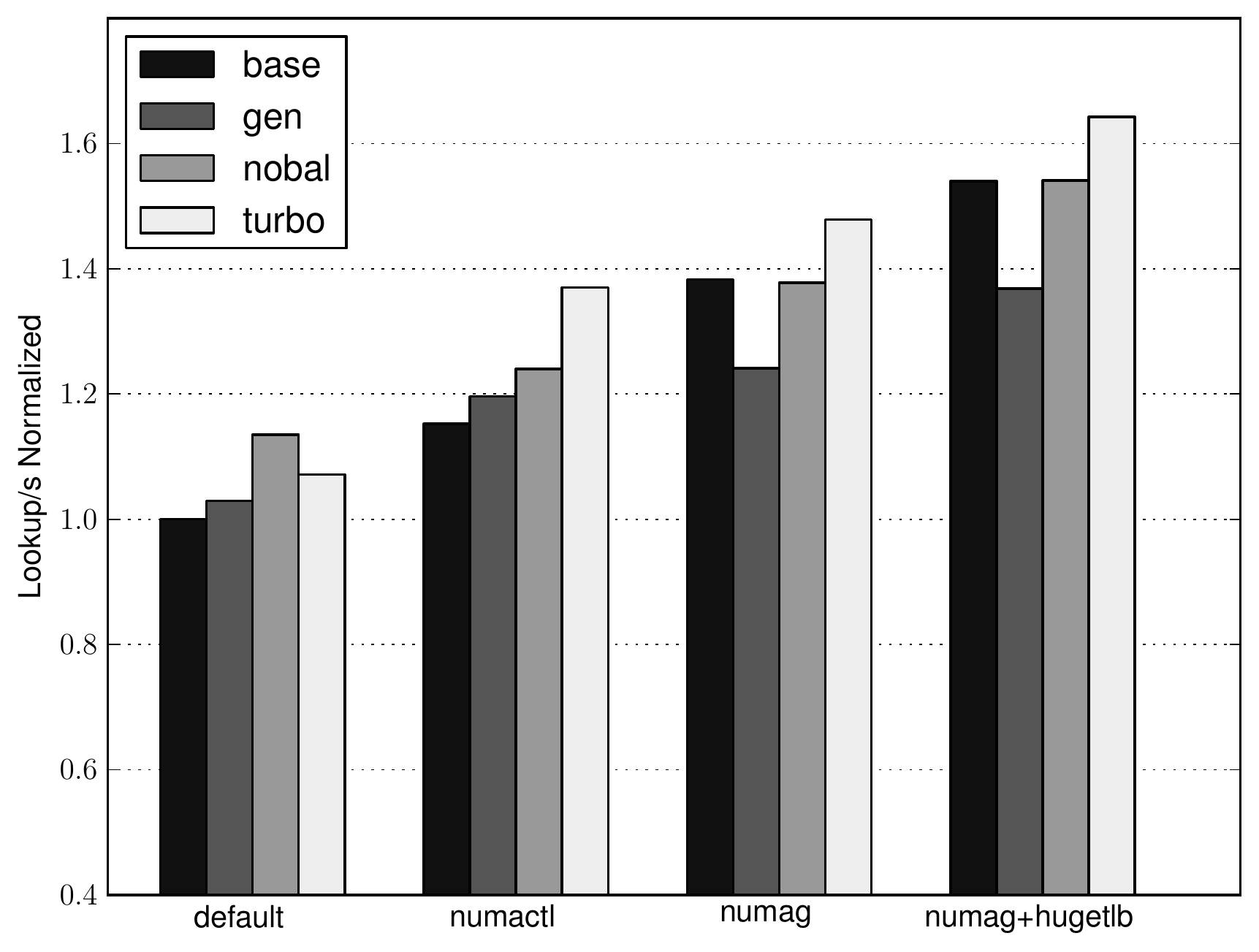}
\caption{Normalized lookup performance at 16 threads}
\label{fig:optxsnorm}
\end{figure}

\begin{figure}
\centering
  \includegraphics[width=.6\columnwidth]{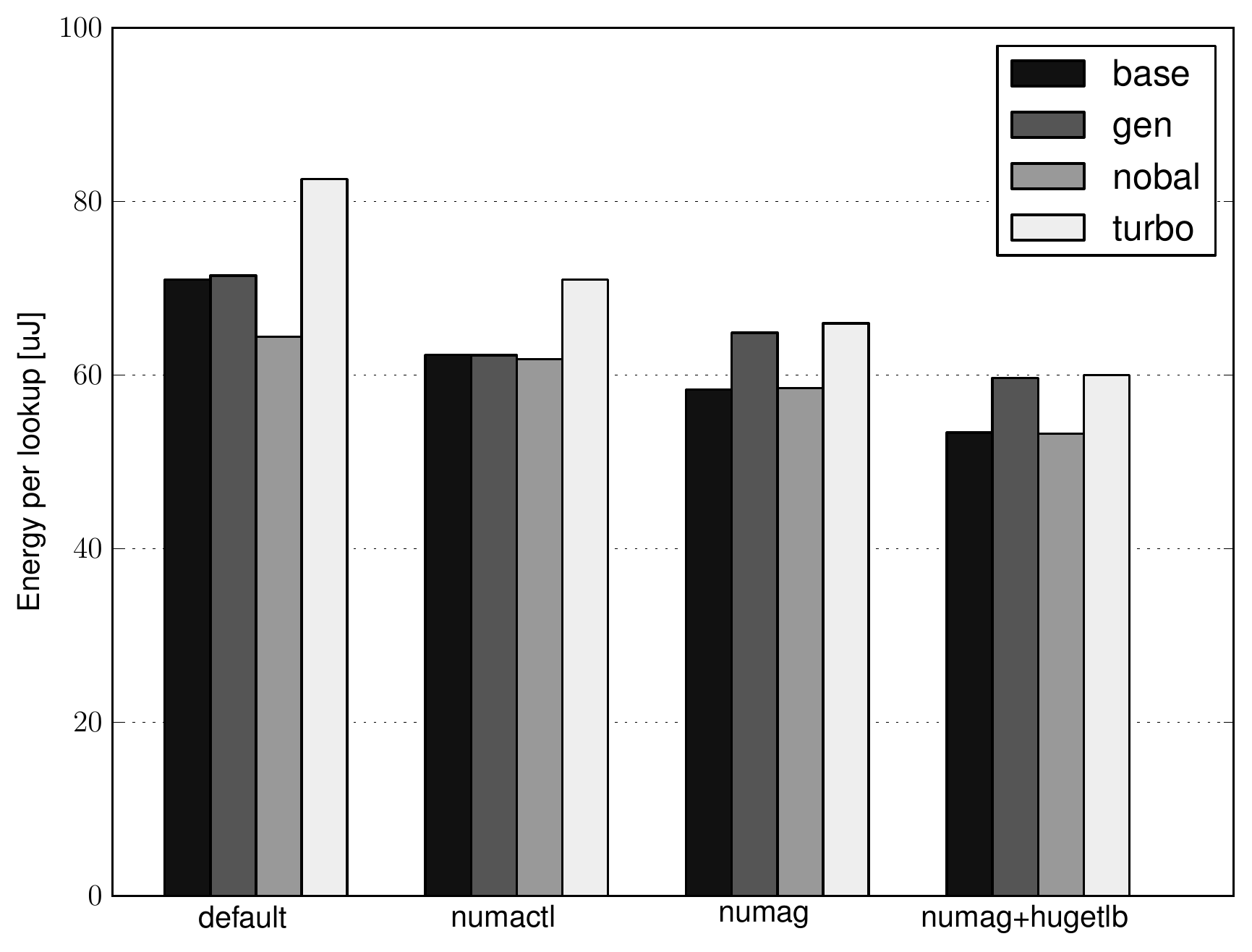}
  \caption{Energy consumption per lookup at 16 threads}
\label{fig:optxsenergy}
\end{figure}

In addition to the memory optimizations,
we explore the running modes described in Table~\ref{tbl:opts}.
Each running mode
is described below.  Figure~\ref{fig:optxsnorm} compares the
normalized lookup performance among the memory optimizations on the
four different running modes. In a similar manner,
Figure~\ref{fig:optxsenergy} compares the energy consumption per
lookup.  

Disabling the kernel-level, automatic NUMA-aware memory placement
(nobal) improves ``default'' and ``numactl,'' but it
has no effect on ``numactl.''
We need to investigate this
behavior further, but the physical pages associated with the nuclides grids may
migrate back and forth between socket0 and socket1. The running mode
``gen'' has a negative impact on ``numag,'' presumably because
the unionized energy grids are not interleaved properly.  We observe
that ``turboboost'' is always effective. However, ``turboboost''
leverages the thermal head room, so the performance may vary.  In
terms of the total energy consumption (CPU and DRAM), the most optimized version consumes
25\% less energy than does the nonoptimized version in the base running mode.


\begin{table} 
\caption{Running Modes}
\label{tbl:opts}
\footnotesize
\centering
\begin{tabular}{|l|l|l|l|}
\hline
\texttt{Label}  & \textbf{Initialization} & \textbf{NUMA Balancing} & \textbf{Frequency}  \\ \hline \hline
base            & file                    & enable          & 2.6\,GHz \\ \hline
gen             & runtime                 & enable          & 2.6\,GHz \\ \hline
nobal           & file                    & disable         & 2.6\,GHz \\ \hline
turbo           & file                    & enable          & boost (up to 3.3\,GHz) \\ \hline
\end{tabular}
\end{table}

\subsubsection{Initialization:}


In the table, the column ``Initialization'' refers to 
how XSBench initializes its data set. The ``file'' mode is
described in Section~\ref{sec:scalability} and
Section~\ref{sec:osnuma}. XSBench reads a pre-computed data set from a
file into the memory buffer allocated during its OpenMP serial region
(by its master thread), which creates an unbalanced data placement.
Initializing its data set in the ``runtime'' mode takes XSBench longer than initializing in the ``file'' mode
because initializing the unionized energy grid is
expensive; the computational cost of the initialization is $n^2 m
\log m$, where $n$ is the total number of isotopes and $m$ is the total
number of gridpoints.  With the default XSBench configuration, initialization
takes approximately 300 seconds on our Sandy Bridge testbed with one
thread. In order to amortize the initialization cost, the OpenMP parallel for
loop is used. With the first-touch policy, the physical memory
allocation is likely to be interleaved in some way.


\subsubsection{NUMA Balancing:}

This refers to the automatic NUMA-aware memory placement kernel option
described in Section~\ref{sec:osnuma}.  The running mode ``nobal''
disables this kernel option, otherwise it is enabled.

\subsubsection{Frequency:}

This refers to the processor frequency configuration.  Aside from the
running mode ``turbo,'' the frequency is set to 2.6\,GHz, which is the
maximum user-configurable frequency on our testbed.  With
``turbo,'' the processor can increase the frequency for a heavy
workload, depending on the current thermal headroom.  Since the
energy consumption depends on temperature and since turboboot is a
thermal-aware mechanism, we cool the processor down below 40 degrees Celsius
before starting XSBench each time in this study.



\section{Related Work}
\label{sec:relatedwork}

Since the concept of NUMA has been around for a while, there are
numerous studies on NUMA architecture and NUMA related memory
management
techniques~\cite{stenstrom92-coma,Bolosky:1989:SBE:74851.74854,larowe89-numa,bolosky91-numa}.
The advent of higher core-count multicore processors and modern
interconnects poses a new, interesting challenge to from algorithms to
system software designs.  Majo et
al.~\cite{Majo:2011:MSP:1987816.1987832} analyzed the memory
controller behavior of the Intel Xeon 5520 processor. They also
developed a model to characterize the memory bandwidth.  They found
that maximizing data locality does not improve the performance and
suggested that allocating data on a remote processor may benefit
applications.  Li et al.~\cite{LiPMRL13} optimize a data shuffling
algorithm for NUMA, considering modern NUMA architecture. They
delineate the bandwidth and latency of a 4-socket Nehalem-EX system
and present the problem of data shuffling in NUMA.  They showed that
the optimized version is three time faster than its naive version. 

Furthermore the scalability of multi-threaded, shared-memory
programming languages or APIs such as OpenMP is also becoming a major
issue on a multicore, NUMA system since those languages or APIs are
originally designed, based on the assumption of uniform memory access
(UMA). Many studies have been conducted to extend shared memory
programming languages to NUMA.  Broquedis et
al.~\cite{BroquedisFGWN10} combined NUMA-aware memory manager with
their runtime system to enable dynamic load distribution, utilizing
the information from the application structure and hardware topology.
Olivier et al.\cite{Olivier:2012:OTS:2237840.2237846} propose a
hierarchical scheduling strategy to improve the performance. They
successfully demonstrated several benchmarks to successfully scale to
192 CPUs of an SGI Altix with their strategy.  

In this study we detail the internal design of the Linux kernel as
well as the algorithm because we believe the increasingly complex
system software layer tend to wider a gap between the hardware and the
application.  In addition we also investigate on the influences of
running modes and the energy consumption.






\section{Conclusion}
\label{sec:conclusion}

The multicore scalability of OpenMP programs on NUMA is becoming a big
issue.  We explain what causes the multicore scalability, presenting
details of both XSBench and the Linux kernel.  We demonstrate that
precise control of the physical memory allocation is important on
NUMA.  We find that a simple technique like duplication can be very
effective. Using the technique we present, we realize a significant
improvement: the scaling efficiency is improved from a nonoptimized
70\% to an optimized 95\% , and the optimized version consumes 25\%
less energy than does the nonoptimized version.
The lookup performance
of XSBench is also improved from 2.1 million Lookups/s to 3.2, with
minimum code modification, which is a considerable improvement.
We also note that both turboboost and
hyper-threading are effective with the memory optimization.
We plan to evaluate the optimization
technique on a four-way or higher NUMA system. Our interests are the
energy and the operating system scalability as well as the scalability
on a bigger NUMA system.


\subsection*{Acknowledgments}

The submitted manuscript has been created by UChicago Argonne, LLC,
Operator of Argonne National Laboratory (``Argonne'').  Argonne, a
U.S. Department of Energy Office of Science laboratory, is operated
under Contract No. DE-AC02-06CH11357.  The U.S. Government retains for
itself, and others acting on its behalf, a paid-up nonexclusive,
irrevocable worldwide license in said article to reproduce, prepare
derivative works, distribute copies to the public, and perform
publicly and display publicly, by or on behalf of the Government.




\bibliographystyle{splncs}
\bibliography{bib/mc,bib/misc,bib/numa,bib/os}


\end{document}